\documentclass[aps,prl,manuscript,showpacs]{revtex4}%
\usepackage{graphicx}%
\usepackage{amsmath}%
\usepackage{amsfonts}%
\usepackage{amssymb}
\begin{document}
\title{Water on Pt(111): the importance of proton disorder}
\author{Luigi Delle Site}
\affiliation{Max-Planck-Institut f\"ur Polymerforschung, Ackermannweg 10, D-55128 Mainz, Germany}
\author{Luca M. Ghiringhelli}
\affiliation{Max-Planck-Institut f\"ur Polymerforschung, Ackermannweg 10, D-55128 Mainz, Germany}
\author{Oliviero Andreussi}
\affiliation{Scuola Normale Superiore, P.$^{\text{zza}}$ dei Cavalieri 7, 56100 Pisa, Italy}
\author{Davide Donadio}
\affiliation{Computational Science, Department of Chemistry and Applied Biosciences ETH
Zurich, USI-Campus, via Giuseppe Buffi 13, CH-6900 Lugano, Switzerland}
\author{Michele Parrinello}
\affiliation{Computational Science, Department of Chemistry and Applied Biosciences ETH
Zurich, USI-Campus, via Giuseppe Buffi 13, CH-6900 Lugano, Switzerland}
\affiliation{Scuola Normale Superiore, P.$^{\text{zza}}$ dei Cavalieri 7, 56100 Pisa, Italy}

\date{\today}

\begin{abstract}
The structure of a water adlayer on Pt(111) surface is investigated by
extensive first principle calculations. Only allowing for proton disorder
the ground state energy can be found. This results from an interplay between water/metal
chemical bonding and the hydrogen bonding of the water network. The resulting short O-Pt
distance accounts for experimental evidences. The novelty of these results shed
a new light on relevant aspects of water-metal interaction.

\end{abstract}

\maketitle

The investigation of the interaction of water with solid metal
surfaces is of extremely high interest in surface chemistry,
catalysis and many other technological applications and is being
carried out extensively by means of experiments and theoretical
calculations\cite{reviewTaylor,reviewMichaelides}. In particular,
much attention has been devoted to the initial stages of the
wetting of transition metal surfaces
\cite{michaelides03,sebastiani05,meng04}, which occurs through the
adsorption of monomers or small clusters, and to the structure and
stability of very thin water overlayers on
Ru(1000)~\cite{feibelman02}, Rh(111)~\cite{feibelman03},
Pt(111)~\cite{meng02,ogasawara02} and Cu(110)~\cite{ren06}. In
spite of these important efforts, the determination of the stable
structure of water overlayers is in several cases still
controversial, as experiments cannot provide conclusive
information at the atomistic level. On metals with low proton
affinity, as in the cases of Pt, Ni and Cu, water arranges in an
undissociated hexagonal pattern~\cite{doering82}, while on
Ru(1000) a half-dissociated monolayer structure has been
demonstrated to be energetically favored
\cite{feibelman02,menzel94}.

The interaction of water monomers and
small clusters on Pt(111) is by far the best
studied~\cite{morgen96,michaelides03}. Water is adsorbed at the
top site and the bond with the substrate is mainly due to the
interaction of the unoccupied $5d$ states of Pt atoms with the
$1b_{1}$ occupied molecular state of water. Nevertheless, the
periodic adlayer can assume three different bilayer arrangements,
namely $(\sqrt{39}\times\sqrt{39})R16.1^{o}$ (RT39),
$(\sqrt{37}\times\sqrt{37})R25.3^{o}$ (RT37) and
$(\sqrt{3}\times\sqrt {3})R30^{o}$
(RT3)~\cite{haq02,glebov97,firment79}, with competing adsorption
energies \cite{meng04}.
However {\sl ab initio} calculations showed that the RT39 is the most energetically
favored structure~\cite{feibelmancomment,meng04}, the RT3 has been 
identified in low energy electron diffraction~\cite{jacobi01} and 
STM experiments~\cite{morgenstern97}. In addition, the difference in the adsorption energy
between the RT3, RT37 and RT39 structures of the wetting layer vanishes
to zero as the water coverage increases \cite{meng04}.
Partially dissociated (OH$+$H$_{2}$O)
structures have only been observed in connection to water
production reaction \cite{clay04} and theoretically studied
\cite{karlberg03,karlberg04}.
Given this, it is clear that one structure prevails on another as a 
function of the kinetics and the thermodynamic conditions of the 
experiments.

While most of the experimental
evidences are well described by the remarkable wealth of
theoretical results in literature, recent XAS\ and XPS results
indicate a water layer distance to the substrate (2.4 \AA) by far
shorter than the theoretically predicted one \cite{ogasawara02}.
Indeed, the previously reported theoretical studies predict that
the most stable water configurations is a bilayer, where the
shortest Pt-O distance (2.8 \AA) is the result of a weak
water-platinum bond. In this configuration the water network is strongly
bound and it is left essentially unchanged by the interaction with
the metal substrate.
\begin{figure}
[b!]
\begin{center}
\includegraphics[width=0.54\columnwidth,clip]{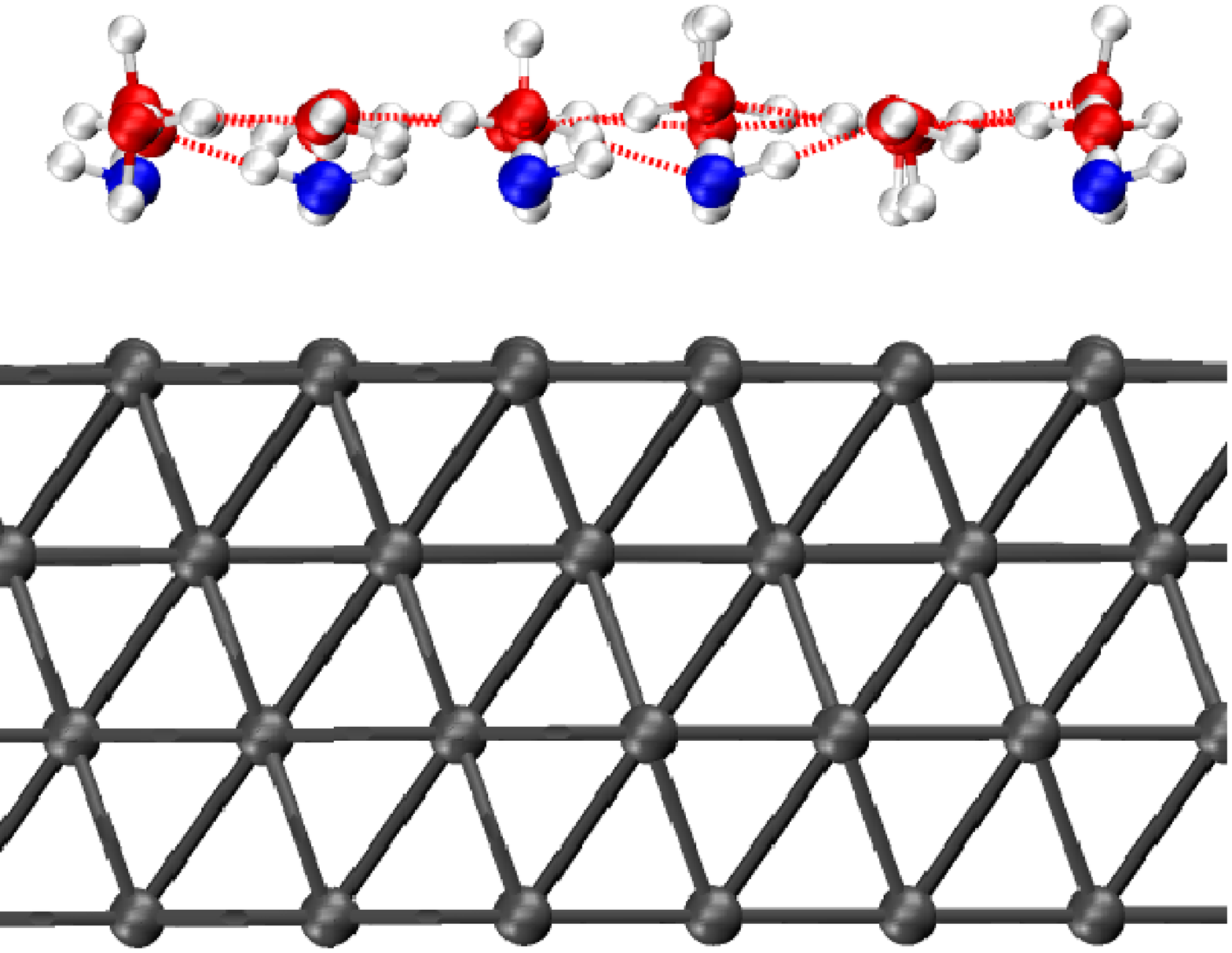}
\includegraphics[width=0.44\columnwidth,clip]{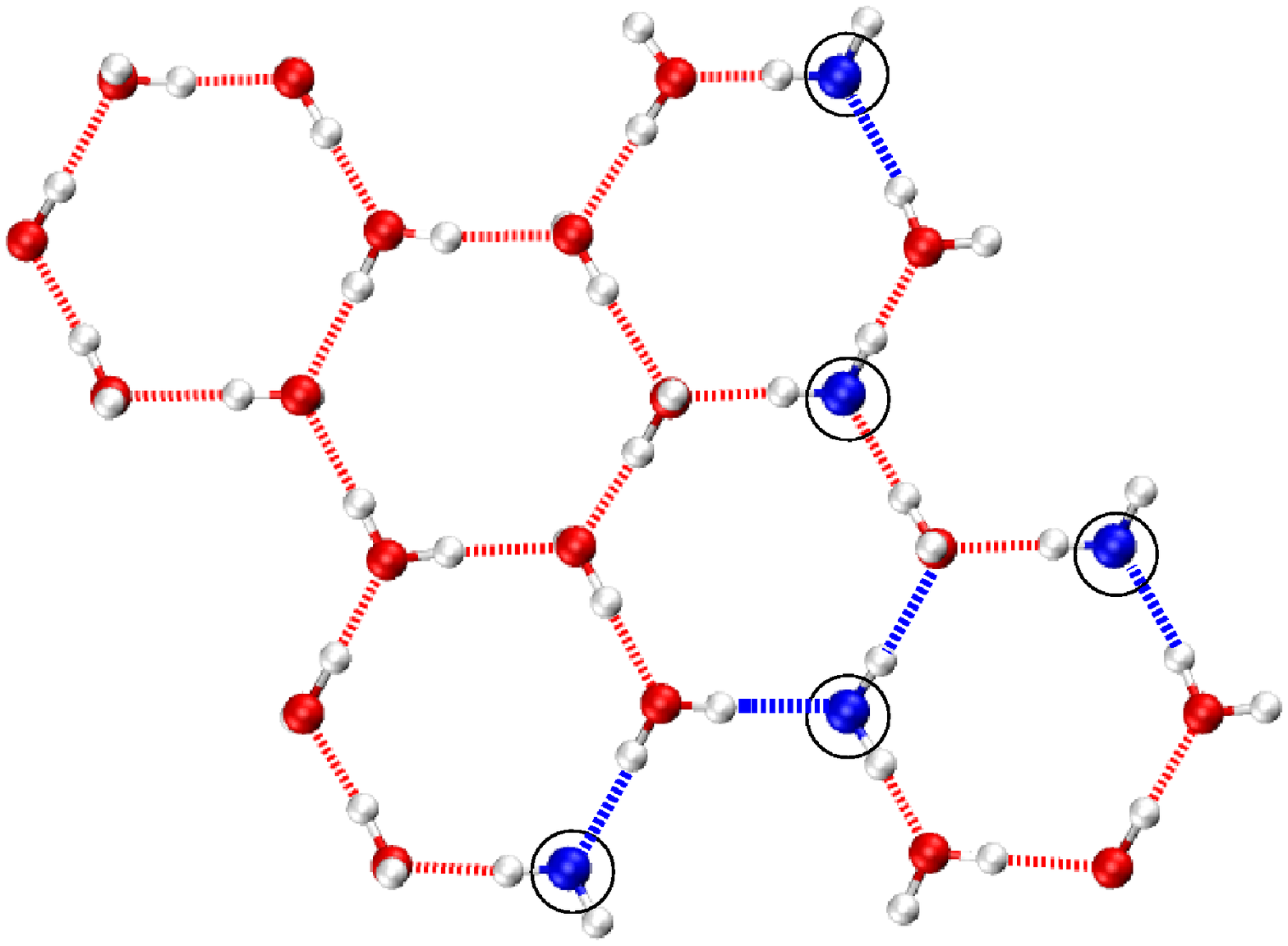}
\caption{(Side and top views of the proton disordered H$_{mix}$
configuration. The five Pt atoms bonded to the oxygens are raised
by 0.1 \AA~with respect to the average $z$-position of the
remaining 31 first layer Pt atoms. The five bonded oxygen atoms
(circled in the top view) stay at 2.33 \AA~from the closest Pt
atoms, while the rest of the water network lays 3.1-3.4A far from
the surface. Hydrogen bonds are drown in red (blue) when
the O-O distance is smaller (greater) than 3.0 \AA~(see text).}%
\label{fig:structures}%
\end{center}
\end{figure}
This discrepancy between experimental evidence and theoretical
results is still an open problem and in this Letter we address
this issue via a computational approach. We use a large unit cell,
since it allows for a less biased search of the lowest energy structure as
underlined in Ref.\cite{vassilev05}, and investigate the possible
role of proton disorder which is known to be essential to
stabilize not only bulk crystals~\footnote{The most common
polymorphs of ice are indeed proton disordered}, but also
crystalline films~\cite{oliviero06}; moreover it has been already
supposed to play an important role in determining the wetting
structures on transition metals
surfaces~\cite{clay04,feibelman03}.We show that for this kind of
system the proton disorder is essential to understand the ground
state structure. In addition, for the fully proton disordered
system, some of the water molecules form a strong bond with the
substrate leading to a short bonding distance (2.3 \AA). The
formation of these strong bonds leads to a weakening of the
hydrogen-bond network and the remaining molecules move to a
larger distance (3.1-3.5 \AA). The resulting picture agrees with
the experimental findings and allows for an interpretation of the
adsorption mechanism.

We use the DFT based finite-electronic temperature method (FEMD)
\cite{alavi94} implemented in the plane-wave based CPMD code
\cite{CPMD}. In this method the electron density and the
Hellman-Feynman forces are determined via a subspace
diagonalization of the high temperature electron density matrix.
The subspace is expanded in a plane-wave basis set that in our
set-up is cutoff at 60 Ry, a value which, according to our
previous work \cite{sebastiani05}, is sufficient to describe
satisfactorily the molecule/metal interaction. We use
pseudopotentials generated according to Troullier-Martins
\cite{tmpp} scheme; all the pseudopotentials were accurately
tested to reproduce bulk and surface properties of the metal, and
correct structural properties for the molecule. We use the PBE
\cite{PBE96} generalized gradient corrected functional. The
prototype system consists of a (111) surface of platinum
represented by 3 and 4 layers (as explained later on regarding the
geometry optimization procedure) and a layer of water composed of
24 molecules. We employ a 6$\times$6 supercell of hexagonal
symmetry with the cell dimension in the direction perpendicular to
the surface equal to 25.5 \AA ~so that the thickness of the vacuum
between the water molecules and the bottom layer of the image slab
of Pt(111) is equal to about 15 \AA; due to the extended size of
the box, we performed the calculations at the $\Gamma$-point
approximation. Geometry optimizations, using the BFGS algorithm,
were done first with only 3 layers until the maximum component of
the ionic forces was below $2\times10^{-3}$ atomic units, then we
add a fourth layer and further optimize the structure until the
convergence of the root-mean square force was below $10^{-3}$
atomic units, the maximum component of the ionic force was at
least below $3\times10^{-3}$ and the energy changes were below
$0.01$ eV, i.e. about $10^{-4}$ eV per molecule into the
adsorption energy of water.

Five different proton arrangements are considered: a fully
proton-ordered configuration (H$_{down}^{ord}$), similar to the
ones already studied in literature, two in-plane proton-disordered
structures (H$_{up}$ and H$_{down} $) and two fully proton
disordered structures, one dissociated (H$_{diss}$) and one
undissociated (H$_{mix}$). Following the convention of previous
works ~\cite{michaelides04} we define H$_{up}$ and H$_{down}$ as
the configuration where the hydrogen atoms not participating to
hydrogen-bonds point toward the vacuum or metal surface,
respectively. The initial configuration of the water film is the
one of an ideal (0001) bilayer of hexagonal ice stretched in the
$xy$ plane so to match the surface lattice parameter of the metal.
In the case of Pt this mismatch is $\sim 8\%$. The proton
disordered configurations have been generated by a Montecarlo
method \cite{buch98}, so to minimize either the global or the
in-plane dipole moment. Obviously the $z$ component of the dipole
moment can be set to zero only if there is no constraint on the
number of protons pointing toward the metal surface: in this case
we have the so called H$_{mix}$ film. In the initial
configurations the water molecules are on the top sites of the
platinum surface, the vertical separation between the oxygen atoms
in the water bilayer is 0.6 \AA, and the lower layer is 2.6 \AA~
apart from the metal slab. In the partially dissociated
configuration (H$_{diss}$), the dissociated protons sit at the
free top sites. In Fig.~\ref{fig:structures} the final geometry of
the H$_{mix}$ system is reported; the geometries of the other
systems are those reported in Ref. \cite{reviewMichaelides}.

The total adsorption energies per molecule are computed as
$E_{ads}=(E_{tot}-E_{Pt}-24E_{H_{2}O})/24$, where $E_{tot}$ is the
energy of the optimized Pt$+$water film system, $E_{Pt}$ the
energy of the bare Pt surface and $E_{H_{2}O}$ the energy of a
water molecule. The values are reported in Tab. \ref{distances}\
and show very good agreement with those of
Ref.~\cite{michaelides04} as for the H$_{down}^{ord}$; moreover
the H$_{mix}$ results the most stable. In order to characterize
the tendency of the hydrogen bonding network to become less tight
as a result of the chemical interaction with the surface, we
monitor the number of ``strong'' hydrogen bonds. An H-bond is
defined as ``strong'' when the O-O distance is smaller than 3.0
\AA. In the initial configurations this distance is 2.98 \AA.

\begin{figure}[t!]
\centering
\includegraphics[width=0.99\columnwidth]{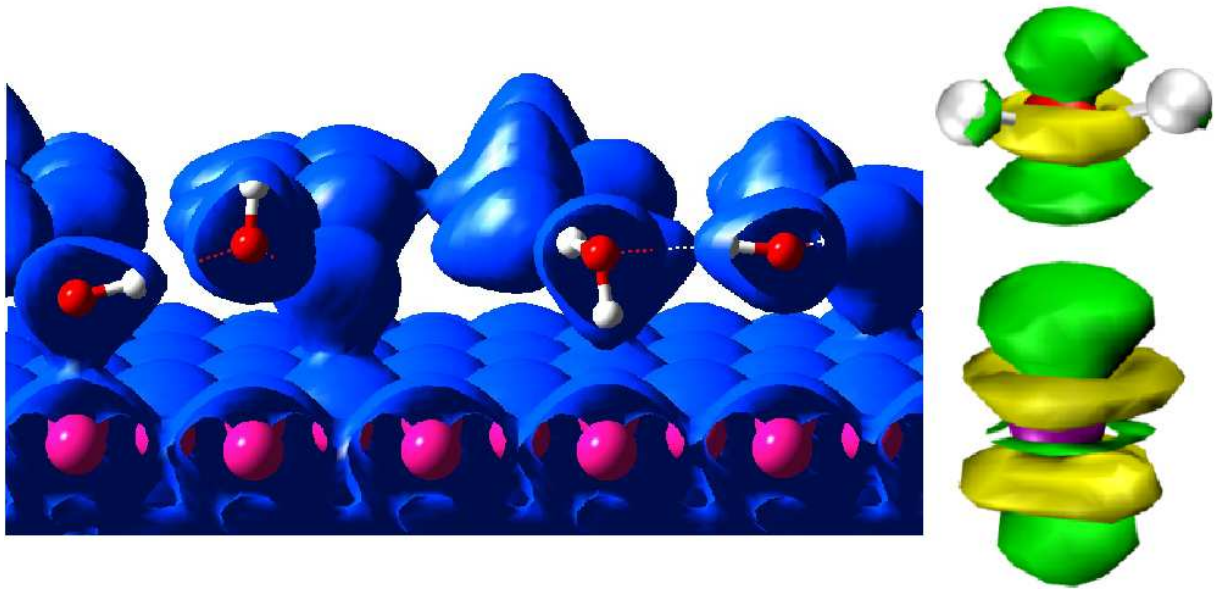}
\caption{Electron density profiles for the H$_{mix}$ case. The
four possible configurations of water molecules are shown in the
left panel (the iso-surface level is 0.270 $e$/\AA$^3$). From left
to right: a flat water molecule in the bottom layer forming a O-Pt
bond, a water molecule with a hydrogen atom pointing to the
vacuum, a water molecule forming a hydrogen bond with the surface
and a flat water molecule in the top layer displaying no
interaction with the surface. Right panel: electron density
difference maps for one of the flat water molecules in the bottom
layer, adsorbed via O-Pt bond. The green (yellow) surface
indicates depletion (accumulation) of electronic density of 0.036
$e$/\AA$^3$. Note that this snapshot compares very well to the
monomer adsorption electron density difference maps shown in
Ref.~\cite{michaelides03}.} \label{fig:chembond}
\end{figure}

In all the undissociated structures, but H$_{mix}$, the H-bond
network stiffens during the optimization. All the H-bonds shorten,
so to recover the ideal O-O distance ($\sim 2.78$ \AA). Since the
underlying metal surface act as a topological confinement, this
process (i.e. O-O shortening) is achieved via a reduction of the
vertical spacing of the bilayer (initially as long as 0.9 \AA).
The water bilayers relaxes at a rather large distance from the Pt
surface, with which it does not interact chemically, as confirmed
by the charge density plot in Fig.\ref{fig:chembond}. In
H$_{diss}$ both the OH groups and the undissociated H$_{2}$O bind
chemically to the metal surface and form a H-bond pattern where
strong (O-O distance: $2.6$ \AA) and very elongated (3.1 \AA)
H-bonds alternate. This picture, confirmed by experiments
\cite{held} and previous calculations \cite{goddard}, suggest a
competition between the chemical O-Pt bonding and the strength of
the H-bond network.

\begin{figure}[t!]
\centering
\includegraphics[width=\columnwidth,clip]{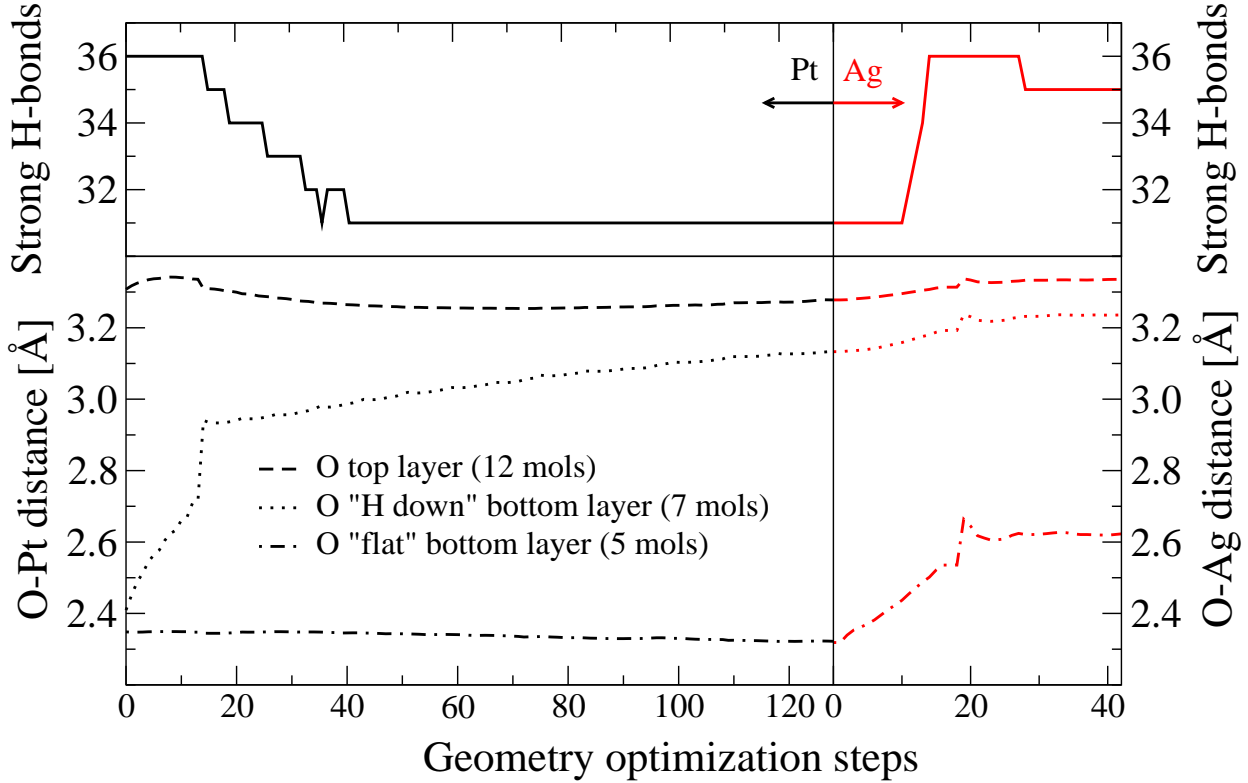}
\caption{Top panels: the number of ``strong'' hydrogen bonds
during the geometry optimization process for the H$_{mix}$ case on
Pt (Ag) surface is reported on the left (right) side. Bottom
panels: Evolution of the distances from the metal surface of the
water molecules for the H$_{mix}$ configuration. In the left
(right) panel we show the distance from the Pt (Ag) surface of the
oxygen of the top water layer (dashed line), of the oxygen of the
H$_{down}$ water molecules in the bottom layer (dotted line), and
(dashed-dotted line) of the ``flat'' water molecules, i.e. those
circled in Fig.~\ref{fig:structures}.} \label{fig:distances}
\end{figure}

This balance determines the adsorption mechanism of the H$_{mix}$
adlayer, which is by far the most stable of the configurations
investigated. Due to the full proton disorder, the system is less
constrained. The optimal balance is found by sacrificing some of
the hydrogen-bond energy of the two dimensional water adlayer
network (top panel of Fig. \ref{fig:distances}) and gaining
chemical energy with the formation of stronger water-metal bonds.
Although the periodicity constraints are relaxed, the water
molecules maintain the initial ordered arrangement in the $xy$
plane, while the distribution of vertical heights is clearly
bimodal (bottom panel of Fig. \ref{fig:distances}). The majority
of water molecules (80$\%$) form a flat overlayer where the Pt-O
distance is distributed between 3.2 and 3.5 \AA, while the flat
water molecules of the bottom layer bind tightly to Pt atoms at a
distance of $\sim$2.3 \AA. The electronic density profiles for the
latter molecules (see the right panel in Fig. \ref{fig:chembond})
shows the formation of chemical bonds with the surface which are
much stronger than in the other structures. Moreover the electron
density difference of the Pt-Me bonding for the H$_{mix}$ case
(left panel in Fig. \ref{fig:chembond}) shows the strong chemical
bond to be almost similar to that of an isolated monomer
\cite{michaelides03}, i.e. the ideal conformation for the
molecular adsorption. As a chemical bond is established between
metal and oxygen atoms the hydrogen bonding of the water adlayer
is weakened, as proven by the elongation of a number of hydrogen
bonds occurring during the optimization of the H$_{mix}$ system
(Fig. \ref{fig:distances}).

In order to provide further evidence to the interplay between Pt-O
chemical bond and the stiffness of the hydrogen bonding network in
the H$_{mix}$ system, we substituted Pt with Ag
\footnote{Computational details are the same as for the other
systems treated here. For Ag we use an LDA pseudopotential
previously tested according to the criteria employed in this work.
The final O-O distances for H$_{mix}$ configuration on Pt, was
scaled according to the lattice constant of Ag, and, as a
consequence, the same is done when counting hydrogen bonds}, which
has a substantially smaller affinity to the adsorption of water
\cite{michaelides03}. In this manner the metal-oxygen chemical
bond is weakened, and performing the geometry optimization we
observe the up-shifting of the previously chemisorbed flat
H$_{2}$O molecules and the restoration of a tight H-bond network
(Fig. \ref{fig:distances}). The calculated adsorption energy of
$-0.46$ eV is in good agreement with the ones reported for proton
ordered systems in Ref. \cite{michaelides04}. This means that for
metal surfaces with low water affinity the issue of proton
disorder is not relevant.

\begin{table}[b!]
\begin{ruledtabular}
\begin{tabular}{lccccc}
& $E_{ads}$ & O-H-Pt & O-Pt & H-O-Pt & OH-Pt \\
\hline
H$^{ord}_{down}$ & $-0.49$ & 3.1 & 2.8 & - & - \\
H$_{down}$ & $-0.50$ & 3.2 & 3.1 & - & - \\
H$_{up}$ & $-0.48$ & - & 2.9 & 3.5 & - \\
H$_{mix}$ & $-0.53$ & 3.2 & 2.3$/$3.2 & 3.3$/$3.5& - \\
H$_{diss}$ & $-0.29$ & - & 2.2 & - & 2.1 \\
\end{tabular}
\end{ruledtabular}
\caption{The adsorption energies (in eV) and the distances between
oxygen and platinum (in \AA) are here reported for the 5 systems
we studied. The labels of the columns have to be interpreted as
follows: O-H-Pt the water molecule has a proton that points toward
the Pt surface, O-Pt the water molecule is flat with respect to
the metal surface, H-O-Pt the water molecule has a proton that
points toward
the vacuum, OH-Pt refers to a dissociated OH group. }%
\label{distances}%
\end{table}%

While providing novel insights about the adsorption mechanism, it
is also remarkable that the H$_{mix}$ on Pt is the only
configuration that accounts for the x-ray absorption/photoemission
spectroscopy measurements in \cite{ogasawara02}, indicating a
relatively strong Pt-O bonding, with a layer of oxygen atoms
2.3--2.4 \AA ~ far from the surface. It must be noticed that, in
none of the other undissociated structures considered in this work
or in the many previously reported
calculations~\cite{ogasawara02,meng02,meng04,michaelides04}, such a
short Pt-O bond has been observed.

In conclusion, we have performed an extensive DFT study of the
structure and bonding of water adlayers on Pt(111) surface, where,
in particular, we have investigated the role of the proton
disorder. We have found that on this particular surface the most
stable adlayer is an undissociated configuration with fully proton
disorder arrangement. The higher stability of the H$_{mix}$
configuration is given by the interplay between the Pt-water
chemical bonding and the capability of the water adlayer to
preserve its hydrogen-bond network intact, though weakened.
We provide the evidence that the main features of the mixed layer/metal
interaction cannot be extrapolated from the study of homogeneously
oriented (all down or all up) layers. In this sense, this study
sheds new light on the nature of the mechanism of interaction of
water on metal substrates and thus opens new paths of experimental
and theoretical investigations into the field.

L.D.S. acknowledges the RZG of the Max Planck Society for the
computational resources, L.M.G. the financial support of the von
Humboldt Foundation. Useful suggestions from K. Kremer are
acknowledged.


\begin{thebibliography}{32}
\expandafter\ifx\csname natexlab\endcsname\relax\def\natexlab#1{#1}\fi
\expandafter\ifx\csname bibnamefont\endcsname\relax
  \def\bibnamefont#1{#1}\fi
\expandafter\ifx\csname bibfnamefont\endcsname\relax
  \def\bibfnamefont#1{#1}\fi
\expandafter\ifx\csname citenamefont\endcsname\relax
  \def\citenamefont#1{#1}\fi
\expandafter\ifx\csname url\endcsname\relax
  \def\url#1{\texttt{#1}}\fi
\expandafter\ifx\csname urlprefix\endcsname\relax\def\urlprefix{URL }\fi
\providecommand{\bibinfo}[2]{#2}
\providecommand{\eprint}[2][]{\url{#2}}

\bibitem[{\citenamefont{Taylor and Neurock}(2006)}]{reviewTaylor}
\bibinfo{author}{\bibfnamefont{C.~D.} \bibnamefont{Taylor}} \bibnamefont{and}
  \bibinfo{author}{\bibfnamefont{M.}~\bibnamefont{Neurock}},
  \bibinfo{journal}{Current Opinion in Solid State and Material Science}
  \textbf{\bibinfo{volume}{9}}, \bibinfo{pages}{49} (\bibinfo{year}{2006}).

\bibitem[{\citenamefont{Michaelides}(2006)}]{reviewMichaelides}
\bibinfo{author}{\bibfnamefont{A.}~\bibnamefont{Michaelides}},
  \bibinfo{journal}{Appl. Phys. A} \textbf{\bibinfo{volume}{85}},
  \bibinfo{pages}{415} (\bibinfo{year}{2006}).

\bibitem[{\citenamefont{Michaelides et~al.}(2003)\citenamefont{Michaelides,
  Ranea, de~Andres, and King}}]{michaelides03}
\bibinfo{author}{\bibfnamefont{A.}~\bibnamefont{Michaelides}},
  \bibinfo{author}{\bibfnamefont{V.~A.} \bibnamefont{Ranea}},
  \bibinfo{author}{\bibfnamefont{P.~L.} \bibnamefont{de~Andres}},
  \bibnamefont{and} \bibinfo{author}{\bibfnamefont{D.~A.} \bibnamefont{King}},
  \bibinfo{journal}{Phys. Rev. Lett.} \textbf{\bibinfo{volume}{90}},
  \bibinfo{pages}{216102} (\bibinfo{year}{2003}).

\bibitem[{\citenamefont{Sebastiani and Delle~Site}(2005)}]{sebastiani05}
\bibinfo{author}{\bibfnamefont{D.}~\bibnamefont{Sebastiani}} \bibnamefont{and}
  \bibinfo{author}{\bibfnamefont{L.}~\bibnamefont{Delle~Site}},
  \bibinfo{journal}{J. Chem. Theory Comput.} \textbf{\bibinfo{volume}{1}},
  \bibinfo{pages}{78} (\bibinfo{year}{2005}).

\bibitem[{\citenamefont{Meng et~al.}(2004)\citenamefont{Meng, Wang, and
  Gao}}]{meng04}
\bibinfo{author}{\bibfnamefont{S.}~\bibnamefont{Meng}},
  \bibinfo{author}{\bibfnamefont{E.~G.} \bibnamefont{Wang}}, \bibnamefont{and}
  \bibinfo{author}{\bibfnamefont{S.}~\bibnamefont{Gao}},
  \bibinfo{journal}{Phys. Rev. B} \textbf{\bibinfo{volume}{69}},
  \bibinfo{pages}{195404} (\bibinfo{year}{2004}).

\bibitem[{\citenamefont{Feibelman}(2002)}]{feibelman02}
\bibinfo{author}{\bibfnamefont{P.~J.} \bibnamefont{Feibelman}},
  \bibinfo{journal}{Science} \textbf{\bibinfo{volume}{295}},
  \bibinfo{pages}{99} (\bibinfo{year}{2002}).

\bibitem[{\citenamefont{Feibelman}(2003{\natexlab{a}})}]{feibelman03}
\bibinfo{author}{\bibfnamefont{P.~J.} \bibnamefont{Feibelman}},
  \bibinfo{journal}{Phys. Rev. Lett.} \textbf{\bibinfo{volume}{90}},
  \bibinfo{pages}{186103} (\bibinfo{year}{2003}{\natexlab{a}}).

\bibitem[{\citenamefont{Meng et~al.}(2002)\citenamefont{Meng, Xu, Wang, and
  Gao}}]{meng02}
\bibinfo{author}{\bibfnamefont{S.}~\bibnamefont{Meng}},
  \bibinfo{author}{\bibfnamefont{L.~F.} \bibnamefont{Xu}},
  \bibinfo{author}{\bibfnamefont{E.~G.} \bibnamefont{Wang}}, \bibnamefont{and}
  \bibinfo{author}{\bibfnamefont{S.}~\bibnamefont{Gao}},
  \bibinfo{journal}{Phys. Rev. Lett.} \textbf{\bibinfo{volume}{89}},
  \bibinfo{pages}{176104} (\bibinfo{year}{2002}).

\bibitem[{\citenamefont{Ogasawara et~al.}(2002)\citenamefont{Ogasawara, Brena,
  Nordlund, Nyberg, Pelmenschikov, Petterson, and Nilsson}}]{ogasawara02}
\bibinfo{author}{\bibfnamefont{H.}~\bibnamefont{Ogasawara}},
  \bibinfo{author}{\bibfnamefont{B.}~\bibnamefont{Brena}},
  \bibinfo{author}{\bibfnamefont{D.}~\bibnamefont{Nordlund}},
  \bibinfo{author}{\bibfnamefont{M.}~\bibnamefont{Nyberg}},
  \bibinfo{author}{\bibfnamefont{A.}~\bibnamefont{Pelmenschikov}},
  \bibinfo{author}{\bibfnamefont{L.~G.~M.} \bibnamefont{Pettersson}},
  \bibnamefont{and} \bibinfo{author}{\bibfnamefont{A.}~\bibnamefont{Nilsson}},
  \bibinfo{journal}{Phys. Rev. Lett.} \textbf{\bibinfo{volume}{89}},
  \bibinfo{pages}{276102} (\bibinfo{year}{2002}).

\bibitem[{\citenamefont{Ren and Meng}(2006)}]{ren06}
\bibinfo{author}{\bibfnamefont{J.}~\bibnamefont{Ren}} \bibnamefont{and}
  \bibinfo{author}{\bibfnamefont{S.}~\bibnamefont{Meng}}, \bibinfo{journal}{J.
  Am. Chem. Soc.} \textbf{\bibinfo{volume}{128}}, \bibinfo{pages}{9282}
  (\bibinfo{year}{2006}).

\bibitem[{\citenamefont{Doering and Madey}(1982)}]{doering82}
\bibinfo{author}{\bibfnamefont{D.~L.} \bibnamefont{Doering}} \bibnamefont{and}
  \bibinfo{author}{\bibfnamefont{T.~E.} \bibnamefont{Madey}},
  \bibinfo{journal}{Surf. Sci.} \textbf{\bibinfo{volume}{123}},
  \bibinfo{pages}{305} (\bibinfo{year}{1982}).

\bibitem[{\citenamefont{Held and Menzel}(1994)}]{menzel94}
\bibinfo{author}{\bibfnamefont{G.}~\bibnamefont{Held}} \bibnamefont{and}
  \bibinfo{author}{\bibfnamefont{D.}~\bibnamefont{Menzel}},
  \bibinfo{journal}{Surf. Sci.} \textbf{\bibinfo{volume}{316}},
  \bibinfo{pages}{92} (\bibinfo{year}{1994}).

\bibitem[{\citenamefont{Morgenstern et~al.}(1996)\citenamefont{Morgenstern,
  Michely, and Comsa}}]{morgen96}
\bibinfo{author}{\bibfnamefont{M.}~\bibnamefont{Morgenstern}},
  \bibinfo{author}{\bibfnamefont{T.}~\bibnamefont{Michely}}, \bibnamefont{and}
  \bibinfo{author}{\bibfnamefont{G.}~\bibnamefont{Comsa}},
  \bibinfo{journal}{Phys. Rev. Lett.} \textbf{\bibinfo{volume}{77}},
  \bibinfo{pages}{703} (\bibinfo{year}{1996}).

\bibitem[{\citenamefont{Haq et~al.}(2002)\citenamefont{Haq, Harnett, and
  Hodgson}}]{haq02}
\bibinfo{author}{\bibfnamefont{S.}~\bibnamefont{Haq}},
  \bibinfo{author}{\bibfnamefont{J.}~\bibnamefont{Harnett}}, \bibnamefont{and}
  \bibinfo{author}{\bibfnamefont{A.}~\bibnamefont{Hodgson}},
  \bibinfo{journal}{Surf. Sci.} \textbf{\bibinfo{volume}{505}},
  \bibinfo{pages}{171} (\bibinfo{year}{2002}).

\bibitem[{\citenamefont{Glebov et~al.}(1997)\citenamefont{Glebov, Graham,
  Menzel, and Toennies}}]{glebov97}
\bibinfo{author}{\bibfnamefont{A.}~\bibnamefont{Glebov}},
  \bibinfo{author}{\bibfnamefont{A.~P.} \bibnamefont{Graham}},
  \bibinfo{author}{\bibfnamefont{A.}~\bibnamefont{Menzel}}, \bibnamefont{and}
  \bibinfo{author}{\bibfnamefont{J.~P.} \bibnamefont{Toennies}},
  \bibinfo{journal}{J. Chem. Phys.} \textbf{\bibinfo{volume}{106}},
  \bibinfo{pages}{9382} (\bibinfo{year}{1997}).

\bibitem[{\citenamefont{Firment and Samoraj}(1979)}]{firment79}
\bibinfo{author}{\bibfnamefont{L.~E.} \bibnamefont{Firment}} \bibnamefont{and}
  \bibinfo{author}{\bibfnamefont{G.~A.} \bibnamefont{Samoraj}},
  \bibinfo{journal}{Surf. Sci.} \textbf{\bibinfo{volume}{84}},
  \bibinfo{pages}{275} (\bibinfo{year}{1979}).

\bibitem[{\citenamefont{Feibelman}(2003{\natexlab{b}})}]{feibelmancomment}
\bibinfo{author}{\bibfnamefont{P.~J.} \bibnamefont{Feibelman}},
  \bibinfo{journal}{Phys. Rev. Lett.} \textbf{\bibinfo{volume}{91}},
  \bibinfo{pages}{059601} (\bibinfo{year}{2003}{\natexlab{b}}).

\bibitem[{\citenamefont{Jacobi et~al.}(2001)\citenamefont{Jacobi, Bedurftig,
  Wang, and Ertl}}]{jacobi01}
\bibinfo{author}{\bibfnamefont{K.}~\bibnamefont{Jacobi}},
  \bibinfo{author}{\bibfnamefont{K.}~\bibnamefont{Bedurftig}},
  \bibinfo{author}{\bibfnamefont{Y.}~\bibnamefont{Wang}}, \bibnamefont{and}
  \bibinfo{author}{\bibfnamefont{G.}~\bibnamefont{Ertl}},
  \bibinfo{journal}{Surf. Science} \textbf{\bibinfo{volume}{472}},
  \bibinfo{pages}{9} (\bibinfo{year}{2001}).

\bibitem[{\citenamefont{Morgenstern et~al.}(1997)\citenamefont{Morgenstern,
  Muller, Michely, and Comsa}}]{morgenstern97}
\bibinfo{author}{\bibfnamefont{M.}~\bibnamefont{Morgenstern}},
  \bibinfo{author}{\bibfnamefont{J.}~\bibnamefont{Muller}},
  \bibinfo{author}{\bibfnamefont{T.}~\bibnamefont{Michely}}, \bibnamefont{and}
  \bibinfo{author}{\bibfnamefont{G.}~\bibnamefont{Comsa}}, \bibinfo{journal}{Z.
  Phys. Chem.} \textbf{\bibinfo{volume}{198}}, \bibinfo{pages}{43}
  (\bibinfo{year}{1997}).

\bibitem[{\citenamefont{Clay et~al.}(2004)\citenamefont{Clay, S.Haq, and
  Hodgson}}]{clay04}
\bibinfo{author}{\bibfnamefont{C.}~\bibnamefont{Clay}},
  \bibinfo{author}{\bibnamefont{S.Haq}}, \bibnamefont{and}
  \bibinfo{author}{\bibfnamefont{A.}~\bibnamefont{Hodgson}},
  \bibinfo{journal}{Phys. Rev. Lett.} \textbf{\bibinfo{volume}{92}},
  \bibinfo{pages}{046102} (\bibinfo{year}{2004}).

\bibitem[{\citenamefont{Karlberg et~al.}(2003)\citenamefont{Karlberg, Olsson,
  Persson, and Wahnstroem}}]{karlberg03}
\bibinfo{author}{\bibfnamefont{G.~S.} \bibnamefont{Karlberg}},
  \bibinfo{author}{\bibfnamefont{F.~E.} \bibnamefont{Olsson}},
  \bibinfo{author}{\bibfnamefont{M.}~\bibnamefont{Persson}}, \bibnamefont{and}
  \bibinfo{author}{\bibfnamefont{G.}~\bibnamefont{Wahnstroem}},
  \bibinfo{journal}{J. Chem. Phys.} \textbf{\bibinfo{volume}{119}},
  \bibinfo{pages}{4865} (\bibinfo{year}{2003}).

\bibitem[{\citenamefont{Karlberg and Wahnstroem}(2004)}]{karlberg04}
\bibinfo{author}{\bibfnamefont{G.~S.} \bibnamefont{Karlberg}} \bibnamefont{and}
  \bibinfo{author}{\bibfnamefont{G.}~\bibnamefont{Wahnstroem}},
  \bibinfo{journal}{Phys. Rev. Lett.} \textbf{\bibinfo{volume}{92}},
  \bibinfo{pages}{136103} (\bibinfo{year}{2004}).

\bibitem[{\citenamefont{Vassilev et~al.}(2005)\citenamefont{Vassilev, van
  Santen, and Koper}}]{vassilev05}
\bibinfo{author}{\bibfnamefont{P.}~\bibnamefont{Vassilev}},
  \bibinfo{author}{\bibfnamefont{R.~A.} \bibnamefont{van Santen}},
  \bibnamefont{and} \bibinfo{author}{\bibfnamefont{M.~T.} \bibnamefont{Koper}},
  \bibinfo{journal}{J. Chem. Phys.} \textbf{\bibinfo{volume}{122}},
  \bibinfo{pages}{054701} (\bibinfo{year}{2005}).

\bibitem[{\citenamefont{Andreussi et~al.}(2006)\citenamefont{Andreussi,
  Donadio, Parrinello, and Zewail}}]{oliviero06}
\bibinfo{author}{\bibfnamefont{O.}~\bibnamefont{Andreussi}},
  \bibinfo{author}{\bibfnamefont{D.}~\bibnamefont{Donadio}},
  \bibinfo{author}{\bibfnamefont{M.}~\bibnamefont{Parrinello}},
  \bibnamefont{and} \bibinfo{author}{\bibfnamefont{A.~H.}
  \bibnamefont{Zewail}}, \bibinfo{journal}{Chem. Phys. Lett.}
  \textbf{\bibinfo{volume}{426}}, \bibinfo{pages}{115} (\bibinfo{year}{2006}).

\bibitem[{\citenamefont{Alavi et~al.}(1994)\citenamefont{Alavi, Kohanoff,
  Parrinello, and Frenkel}}]{alavi94}
\bibinfo{author}{\bibfnamefont{A.}~\bibnamefont{Alavi}},
  \bibinfo{author}{\bibfnamefont{J.}~\bibnamefont{Kohanoff}},
  \bibinfo{author}{\bibfnamefont{M.}~\bibnamefont{Parrinello}},
  \bibnamefont{and} \bibinfo{author}{\bibfnamefont{D.}~\bibnamefont{Frenkel}},
  \bibinfo{journal}{Phys. Rev. Lett.} \textbf{\bibinfo{volume}{73}},
  \bibinfo{pages}{2599} (\bibinfo{year}{1994}).

\bibitem[{CPM()}]{CPMD}
\bibinfo{note}{\textsc{CPMD}, version 3.10.5, developed by J. Hutter \textit{et
  al.}, MPI f\"ur Festk\"orperforschung and IBM Zurich Research Laboratory}.

\bibitem[{\citenamefont{Troullier and Martins}(1991)}]{tmpp}
\bibinfo{author}{\bibfnamefont{N.}~\bibnamefont{Troullier}} \bibnamefont{and}
  \bibinfo{author}{\bibfnamefont{J.~L.} \bibnamefont{Martins}},
  \bibinfo{journal}{Phys. Rev. B} \textbf{\bibinfo{volume}{43}},
  \bibinfo{pages}{1993} (\bibinfo{year}{1991}).

\bibitem[{\citenamefont{Perdew et~al.}(1996)\citenamefont{Perdew, Burke, and
  Ernzerhof}}]{PBE96}
\bibinfo{author}{\bibfnamefont{J.~P.} \bibnamefont{Perdew}},
  \bibinfo{author}{\bibfnamefont{K.}~\bibnamefont{Burke}}, \bibnamefont{and}
  \bibinfo{author}{\bibfnamefont{M.}~\bibnamefont{Ernzerhof}},
  \bibinfo{journal}{Phys. Rev. Lett.} \textbf{\bibinfo{volume}{77}},
  \bibinfo{pages}{3865} (\bibinfo{year}{1996}).

\bibitem[{\citenamefont{Michaelides et~al.}(2004)\citenamefont{Michaelides,
  Alavi, and King}}]{michaelides04}
\bibinfo{author}{\bibfnamefont{A.}~\bibnamefont{Michaelides}},
  \bibinfo{author}{\bibfnamefont{A.}~\bibnamefont{Alavi}}, \bibnamefont{and}
  \bibinfo{author}{\bibfnamefont{D.~A.} \bibnamefont{King}},
  \bibinfo{journal}{Phys. Rev. B} \textbf{\bibinfo{volume}{69}},
  \bibinfo{pages}{113404} (\bibinfo{year}{2004}).

\bibitem[{\citenamefont{Buch et~al.}(1998)\citenamefont{Buch, Sandler, and
  Sadlej}}]{buch98}
\bibinfo{author}{\bibfnamefont{V.}~\bibnamefont{Buch}},
  \bibinfo{author}{\bibfnamefont{P.}~\bibnamefont{Sandler}}, \bibnamefont{and}
  \bibinfo{author}{\bibfnamefont{J.}~\bibnamefont{Sadlej}},
  \bibinfo{journal}{J. Phys. Chem. B} \textbf{\bibinfo{volume}{102}},
  \bibinfo{pages}{8641} (\bibinfo{year}{1998}).

\bibitem[{\citenamefont{Held et~al.}(2005)\citenamefont{Held, Clay, Barrett,
  Haq, and Hodgson}}]{held}
\bibinfo{author}{\bibfnamefont{G.}~\bibnamefont{Held}},
  \bibinfo{author}{\bibfnamefont{C.}~\bibnamefont{Clay}},
  \bibinfo{author}{\bibfnamefont{S.~D.} \bibnamefont{Barrett}},
  \bibinfo{author}{\bibfnamefont{S.}~\bibnamefont{Haq}}, \bibnamefont{and}
  \bibinfo{author}{\bibfnamefont{A.}~\bibnamefont{Hodgson}},
  \bibinfo{journal}{J. Chem. Phys.} \textbf{\bibinfo{volume}{123}},
  \bibinfo{pages}{064711} (\bibinfo{year}{2005}).

\bibitem[{\citenamefont{Jacob and Goddard~III}(2004)}]{goddard}
\bibinfo{author}{\bibfnamefont{T.}~\bibnamefont{Jacob}} \bibnamefont{and}
  \bibinfo{author}{\bibfnamefont{W.~A.} \bibnamefont{Goddard~III}},
  \bibinfo{journal}{J. Am. Chem. Soc.} \textbf{\bibinfo{volume}{126}},
  \bibinfo{pages}{9360} (\bibinfo{year}{2004}).

\end{thebibliography}

\end{document}